\documentclass[twocolumn,showpacs]{revtex4}
\usepackage{graphicx}

\newcommand{\vecS}{\mbox{\boldmath $S$}}

\begin{document}
\title{Kosterlitz-Thouless transition in planar spin models with bond dilution} 
\author{Tasrief Surungan}
\email{tasrief@unhas.ac.id}
\affiliation{Department of Physics, Hasanuddin University, Makassar 90245, Indonesia}
\affiliation{Department of Physics, Tokyo Metropolitan University, Hachioji, Tokyo 192-0397, Japan}
\author{Yutaka Okabe}
\email{okabe@phys.metro-u.ac.jp}
\affiliation{Department of Physics, Tokyo Metropolitan University, Hachioji, Tokyo 192-0397, Japan}

\date{\today}

\begin{abstract}
We study the two-dimensional bond-diluted XY and six-state clock models 
by Monte Carlo simulation with cluster spin updates. 
Various concentrations of depleted bonds were simulated, in which 
we found a systematic decrease of the Kosterlitz-Thouless (KT) 
transition temperatures of both XY and six-state clock models 
as the concentration of dilution decreases. 
For the six-state clock model, a second KT transition at lower temperature 
was observed.  The KT transition temperatures as well as 
the decay exponent $\eta$ 
for each concentration of dilution are estimated. It is observed that 
the quasi long range order disappears at 
the concentration of dilution very close to the percolation threshold. 
The decay exponent $\eta$ of the KT transitions calculated 
at each concentration indicates that the universality class 
belongs to the pure XY and clock models, analogous to the expectation 
of the Harris criterion for the irrelevance of randomness 
in the continuous phase transition of systems 
with non-diverging specific heat. 
\end{abstract}


\maketitle

\section{Introduction}

Dilution is an indispensable aspect of both theoretical and 
experimental studies due to the presence of defects and impurities 
in any real material~\cite{stinch,dotsenko}.  
The pioneering work by Harris~\cite{harris} on the effect of dilution 
on the critical behavior of systems with continuous transition 
has stimulated many studies in the field of random systems.  
Based on the Harris criterion, it is predicted that the dilution 
will be relevant (irrelevant) if the specific heat exponent $\alpha$ 
of the pure system is positive (negative), and becomes marginal 
in the case $\alpha=0$, for example, in the two-dimensional (2D) 
Ising system~\cite{tomita}.  

It is well known that the pure 2D XY model cannot have 
a true long range order at any finite temperature 
due to the continuous symmetry $U(1)$~\cite{mermin,hohenberg}.   
However, the system can exist in a quasi long range order (QLRO), 
an intermediate phase which is a topological excitation formed 
by vortex-antivortex pairs~\cite{kosterlitz}.  
The number of vortex-antivortex pairs increases with the temperature 
until the system experiences the Kosterlitz-Thouless (KT) transition.  
This transition corresponds to the unbinding of vortex-antivortex pairs, 
which leads the system to a high-temperature disordered phase.  

The existence of QLRO in the presence of dilution is an interesting topic.  
This is related to the fact that the QLRO is a topological order 
vulnerable to the local defect or perturbation.  
The diluted XY models may have the relevance to the study of 
superconductivity, in particular the interaction between vortices 
and the spatial inhomogeneity due to the impurities. 
However, not so much attention has been given 
to the dilution effect on the KT transition.  

Quite recently, two contradictory results on the 2D site diluted 
XY model were reported~\cite{leonel,berche}. 
By using the Monte Carlo (MC) simulation with the Metropolis algorithm, 
Leonel {\it et al.}~\cite{leonel} showed that the QLRO disappears 
before the concentration of vacant sites reaches 
the percolation threshold $p_c$ of site dilution.   
On the other hand, performing a more extensive MC study 
with the Wolff cluster algorithm~\cite{wolff}, 
Berche {\it et al.}~\cite{berche} suggested 
that the QLRO remains up to dilute concentration very close to 
the percolation threshold $p_c$ 
(for the site dilution on the square lattice $p_c \sim 0.593$).  
With these two inconsistent results, it is worth carrying out 
a detailed study of the dilution problem from a different point of view. 
We here treat a bond-diluted case; it should be made clear 
whether the QLRO remains or not 
up to the percolation threshold of bond dilution. 

The effect of the $q$-fold symmetry-breaking fields on the 2D XY model 
has been paid attention~\cite{Jose,tomita1}.  
Treating clock models, where only discrete angles of the XY spins are allowed, 
is essentially equivalent to probing the $q$-fold symmetry-breaking fields. 
The $U(1)$ symmetry of the XY model is replaced by the discrete $Z_q$ symmetry in the $q$-state clock model.  
It was shown~\cite{Jose} that the 2D $q$-state clock model has 
two phase transitions of KT type at $T_1$ and $T_2$ ($T_1<T_2$) for $q>4$. 
There is an intermediate XY-like QLRO phase between a low-temperature 
ordered phase ($T<T_1$) and a high-temperature disordered phase ($T>T_2$). 
The effect of dilution on the low-temperature ordered phase 
due to the discrete symmetry of the clock model is another interesting 
subject to study. 

In this paper, we study the bond-diluted XY and six-state clock models 
($q=6$) on the square lattice.  We use the Monte Carlo method 
with cluster flip.  For the estimator of the KT transition 
we use the ratio of magnetic correlation functions 
with different distances~\cite{tomita2}.  
In the next section, we describe our model and the detail of 
calculation method.  Then, in Sec.~\ref{dilresult} we shall present 
our results.  Sec.~\ref{conclu} is devoted 
to the summary and concluding remarks.

A part of the preliminary results of the present work was reported 
at the conference, "Statistical Physics of Disordered Systems 
and its Application", 
which was held in July 2004, at Hayama, Japan~\cite{okabe}. 

\section{Model and Simulation Method}

The bond-diluted XY spin model is written with the Hamiltonian 
\begin{equation}\label{ham}
  H = \sum_{\langle ij \rangle} J_{ij} \, \vecS_i \cdot \vecS_j 
    = \sum_{\langle ij \rangle} J_{ij} \cos(\theta_i-\theta_j), 
\end{equation}
where summation is performed over the whole nearest neighbors 
 pairs $\langle ij \rangle$, $\vecS_i$ being a unit planar
 spin vector occupying the $i$-th site
of lattice system (here we deal with the square lattice), 
and $\theta_i$ the angle associated with the $i$-th spin.  
For the six-state clock model, the angle takes discrete values, $\theta_i = 2\pi q/6$ 
with $q=0,\cdots,5$.  The quenched dilution is conveyed by
 the coupling $J_{ij}$ following 
a distribution
 $ P(J_{ij})= p \delta(J_{ij} - J) + (1 - p) \delta(J_{ij})$, 
with $p$ being the concentration of existing bonds, or we can say 
$(1-p)$ is the concentration of the diluted (missing) bonds.

We make use of the canonical sampling MC method with multi-cluster 
spin updates (Swendsen-Wang algorithm~\cite{swendsen}).  
The embedded cluster idea for continuous spins due to Wolff~\cite{wolff} 
is adopted, namely by projecting 
the planar spins into a random axis so that the Kasteleyn-Fortuin~\cite{kaste} 
procedure on Ising spins can be performed.  
Spins with missing bonds due to dilution are not connected 
in forming a cluster. 

We simulated the 2D diluted XY and six-state clock models 
on the square lattice with periodic boundary conditions.  
We treated both models with the linear sizes 
of $L$ = 32, 48, 64, 80, and 96.   
Various bond concentrations, from the pure case $p=1.0$ 
down to $p = 0.55$, were simulated.  
For each concentration of each system size we treated 
many different realizations in order to get better statistics; 
typical number of realizations is 256, except for $p =0.55$ 
where more realizations were taken into account 
to compensate highly sample-dependent results. 
We performed $10^4$ MC steps for the equilibration and 
$4 \times 10^4$ MC steps for the measurement.  
We use the reweighting technique~\cite{ferrenberg} 
to obtain the thermal average at temperatures different from 
those at which actual simulations were made.

\section{Results}
\label{dilresult}

\subsection{Specific heat}

\begin{figure}
\begin{center}
\includegraphics[width=0.9\linewidth]{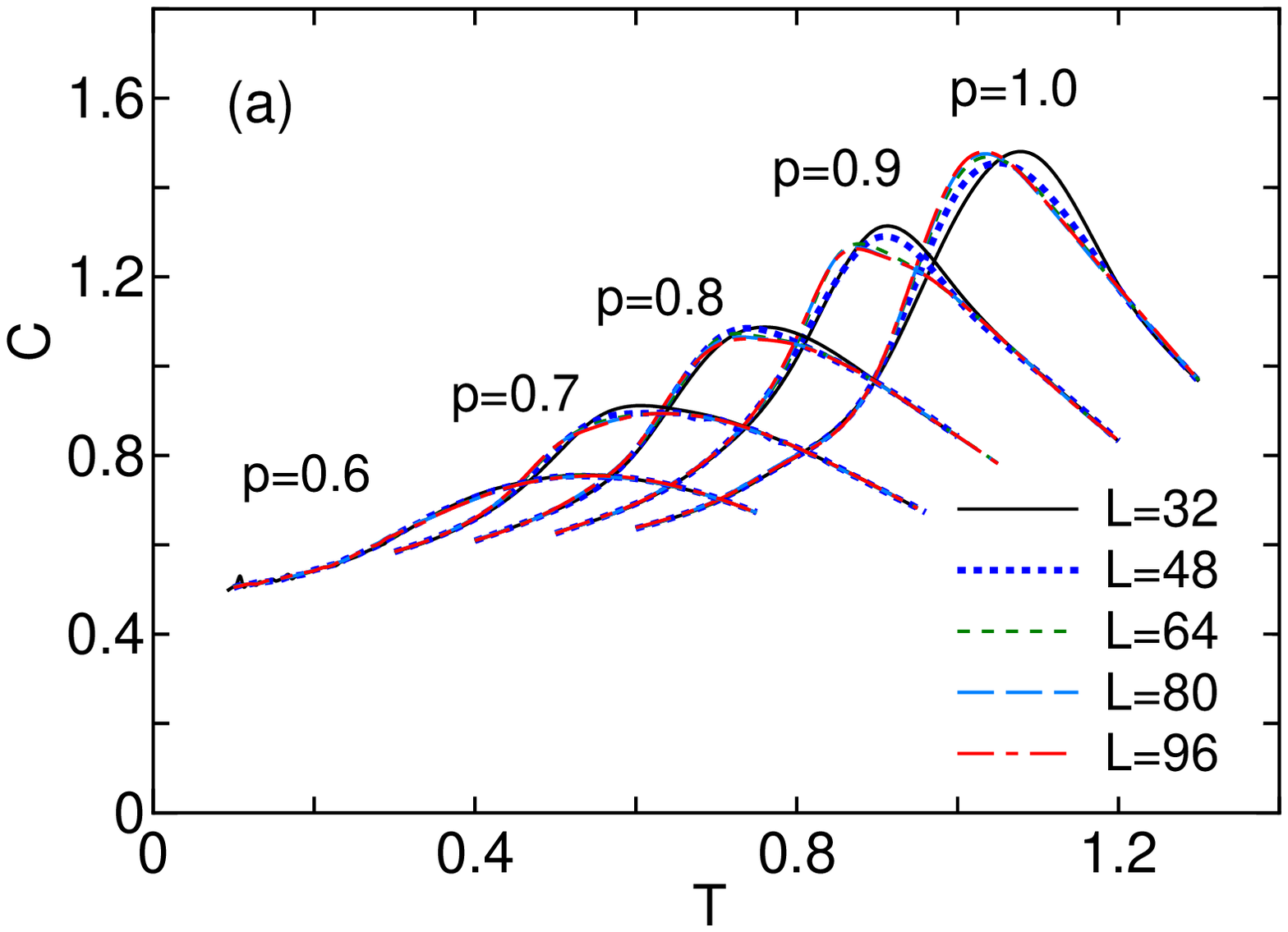}

\vspace{2mm}
\includegraphics[width=0.9\linewidth]{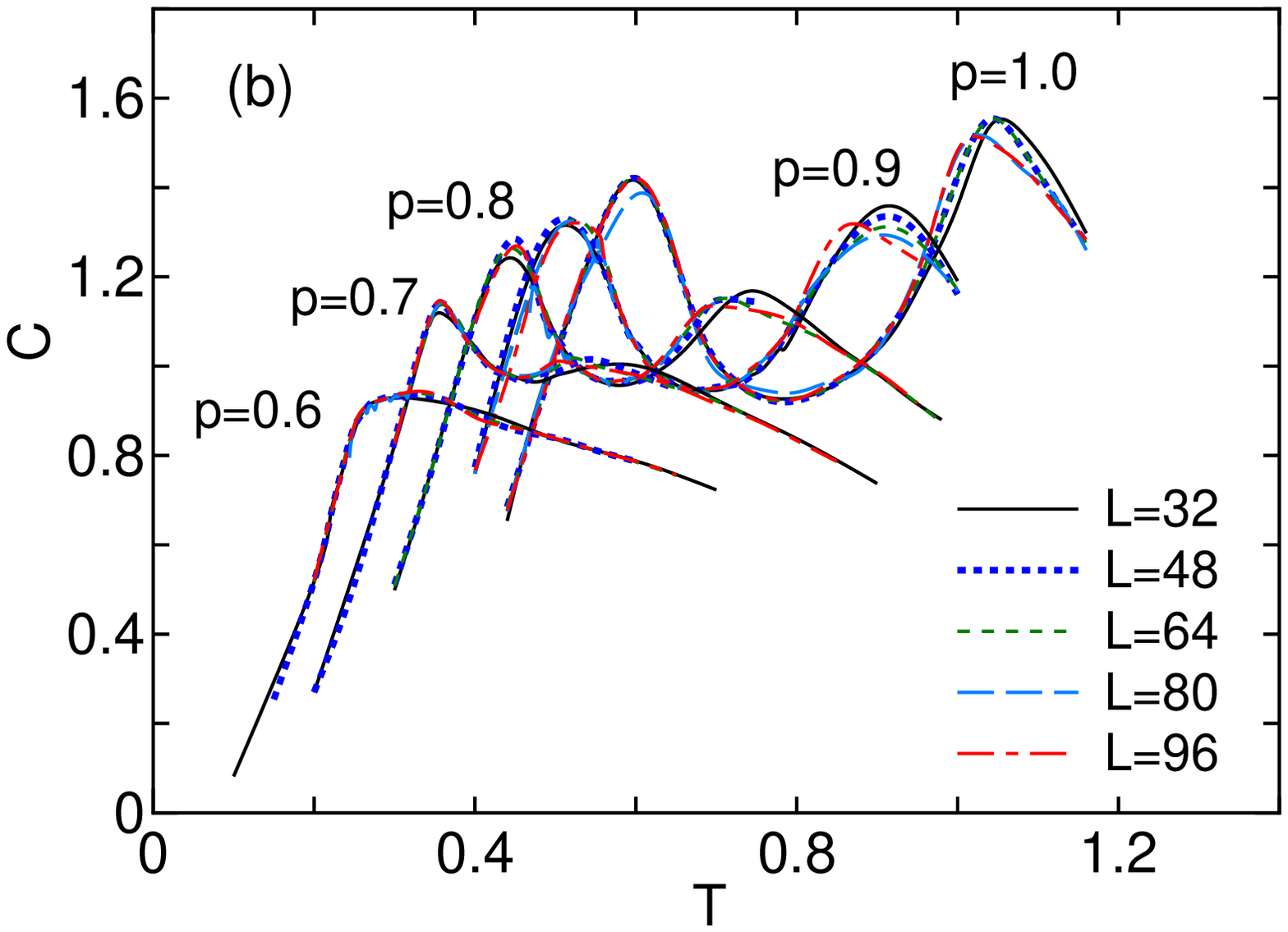}
\caption{Temperature dependence of specific heat of the diluted (a) XY and 
(b) six-state clock models for various bond concentrations 
of system sizes $L$ = 32, 48, 64, 80, and 96. 
}
\label{dilsphtxy} 
\end{center}
\end{figure}

Let us start with presenting the results of the specific heat 
for the diluted XY and six-state clock models on the square lattice. 
The specific heat per spin is defined as follows 
\begin{equation}
 C(T)=\frac{1}{NkT^2}[\langle E^2 \rangle - \langle E \rangle^2], 
\end{equation}
where $k$ is the Boltzmann constant.  The number of spins 
and the total energy are denoted by $N$ and $E$, respectively.

\begin{figure*}
\begin{center}
\includegraphics[width=0.8\linewidth]{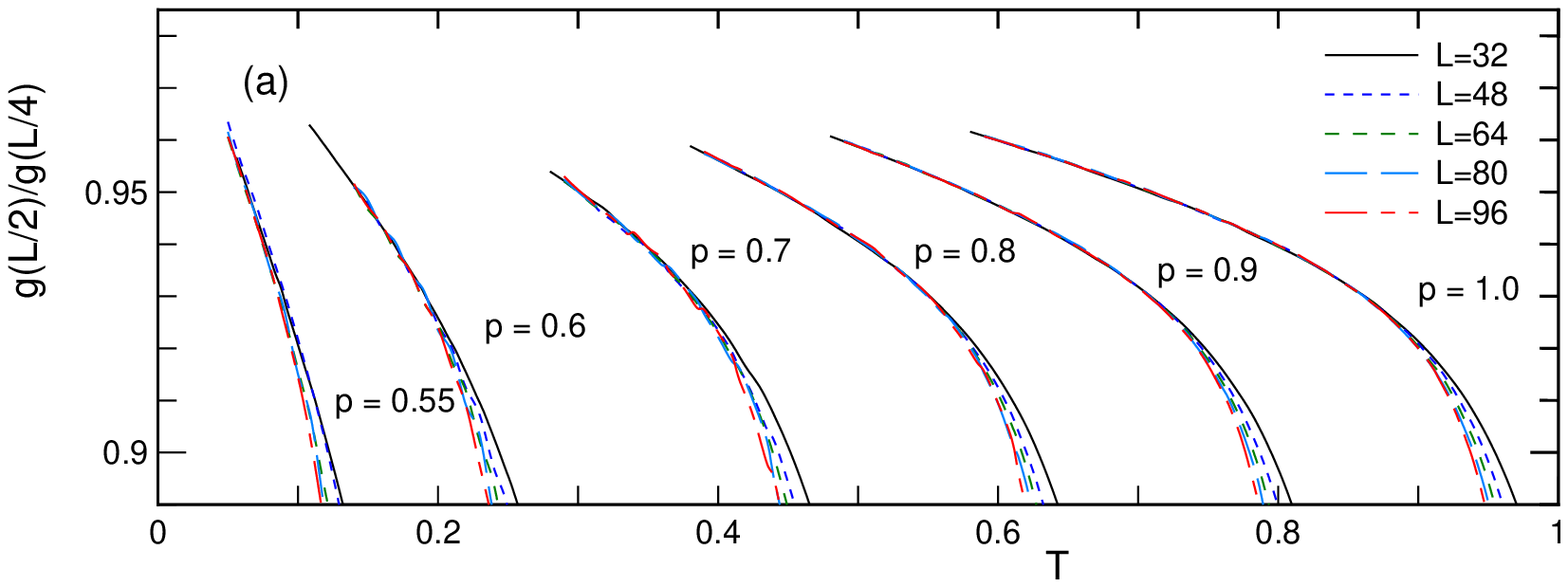}

\vspace{2mm}
\includegraphics[width=0.8\linewidth]{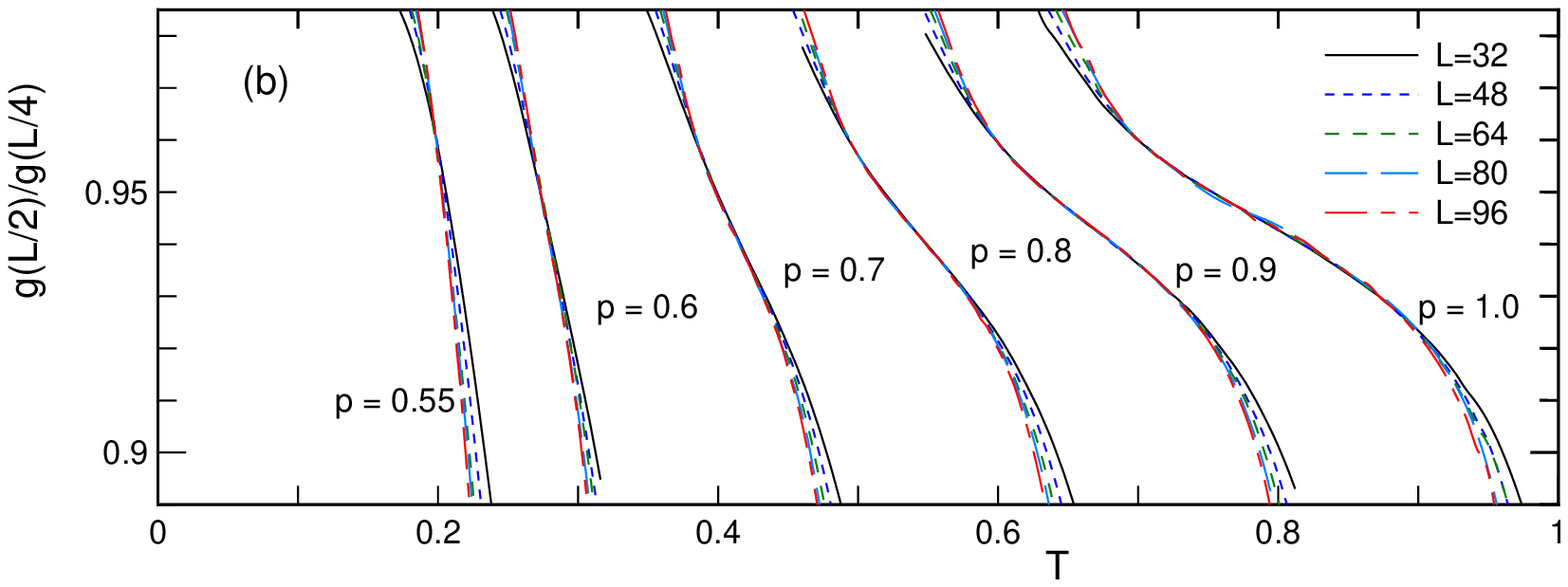}
\caption{Ratio of magnetic correlation functions of the diluted (a) XY 
and (b) six-state clock models 
for various bond concentrations of sizes $L$ = 32, 48, 64, 80, and 96. 
The temperature is represented in units of $J/k$. } 
\label{ratioxy}
\end{center}
\end{figure*}

The temperature dependence of specific heat for various bond concentrations 
are plotted in Fig.~\ref{dilsphtxy} for the diluted XY and 
six-state clock models.  
The temperature is represented in units of $J/k$ from now on.  
The statistical errors are less than the order of the width of curves. 
We see that the size dependence is small, 
which is typical for the KT transition. 
Single finite peaks observed in the diluted XY model may correspond 
to the existence of one KT transition; 
on the other hand there are double peaks for the diluted clock models, 
which signify the existence of two KT transitions. 
In both models, the positions of the peaks gradually shift 
to the lower temperature as we reduce the bond concentration; 
the peaks become relatively flat with the decrease of $p$.  
This indicates that the QLRO in the system gradually fades away.  
However, no abrupt change has been observed; 
the change with $p$ is smooth and continuous. 
More quantitative analysis on the critical behavior shall be 
performed on the magnetic correlation ratio in the following. 

\subsection{Magnetic correlation ratio}

The critical behavior of magnetic ordering  can be investigated 
more precisely from the evaluation of the magnetic  correlation
 function which is defined as the following 
\begin{equation}\label{magchi1}
 g(r) = \langle \vecS_{i} \cdot \vecS_{i+r} \rangle, \label{magchi2}
\end{equation}
where $r$ is the fixed distance between spins.  
Precisely, the distance $r$ is a vector, but we have used 
a simplified notation. 

Consider the ratio of the correlation functions with different distances.  
At the critical point or on the critical line, 
the correlation function $g(r)$ for an infinite system decays as a power of $r$,%
\begin{equation}
  g(r) \sim r^{-(D-2+\eta)}, 
\label{g(r)}
\end{equation}
where $D$ is the spatial dimension and $\eta$ the decay exponent. 
For a finite system in the critical region, 
it can be shown that the ratio of the correlation functions with different 
distances has a finite-size scaling (FSS) form with a single scaling variable,
\begin{equation}
  \frac{g(r,t,L)}{g(r',t,L)} = f(L/\xi),
\label{corr_ratio}
\end{equation}
if we fix two ratios, $r/L$ and $r/r'$, 
where $\xi$ is the correlation length.

Tomita and Okabe~\cite{tomita2} showed that 
this correlation ratio with different distances 
is a very good estimator for the analysis of the second-order phase transition 
as well as for the KT transition.  
The helicity modulus~\cite{minnhagen} and the Binder parameter~\cite{binder} 
are often used in the analysis of the KT transition, but 
the correlation ratio is more efficient in the sense that 
corrections to FSS are smaller~\cite{tomita2}.  It has been 
successfully applied to the study of 
the 2D fully frustrated clock model~\cite{tasrief}. 

In the present work, we consider the ratio of 
magnetic correlation functions setting $r=L/2$ and $r'=L/4$ 
for two distances.  Thus, we evaluate the correlation ratios 
$g(L/2)/g(L/4)$.

\subsubsection{Estimate of KT transition temperatures}

\begin{figure}
\begin{center}
\includegraphics[width=0.9\linewidth]{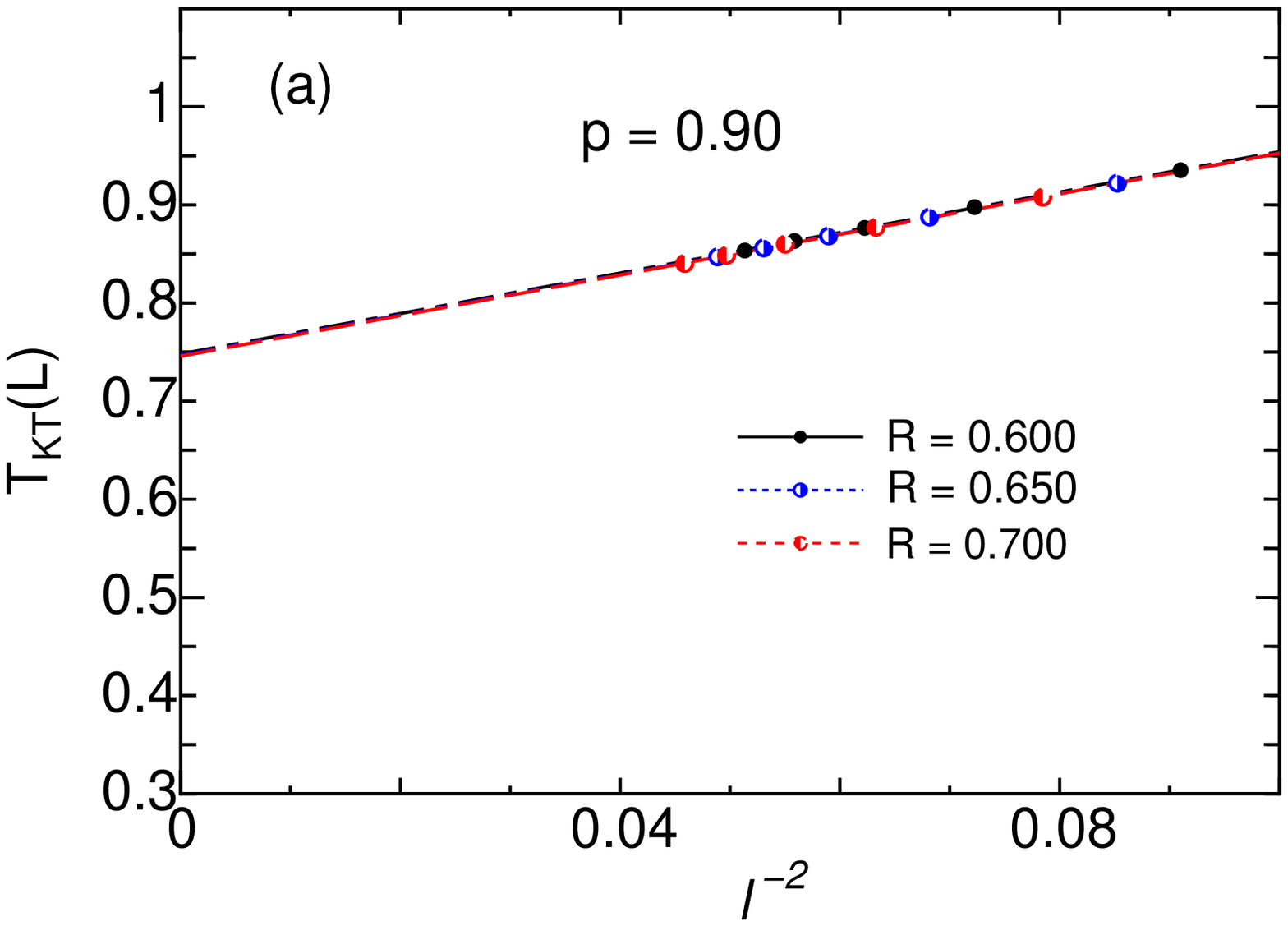}

\vspace{2mm}
\includegraphics[width=0.9\linewidth]{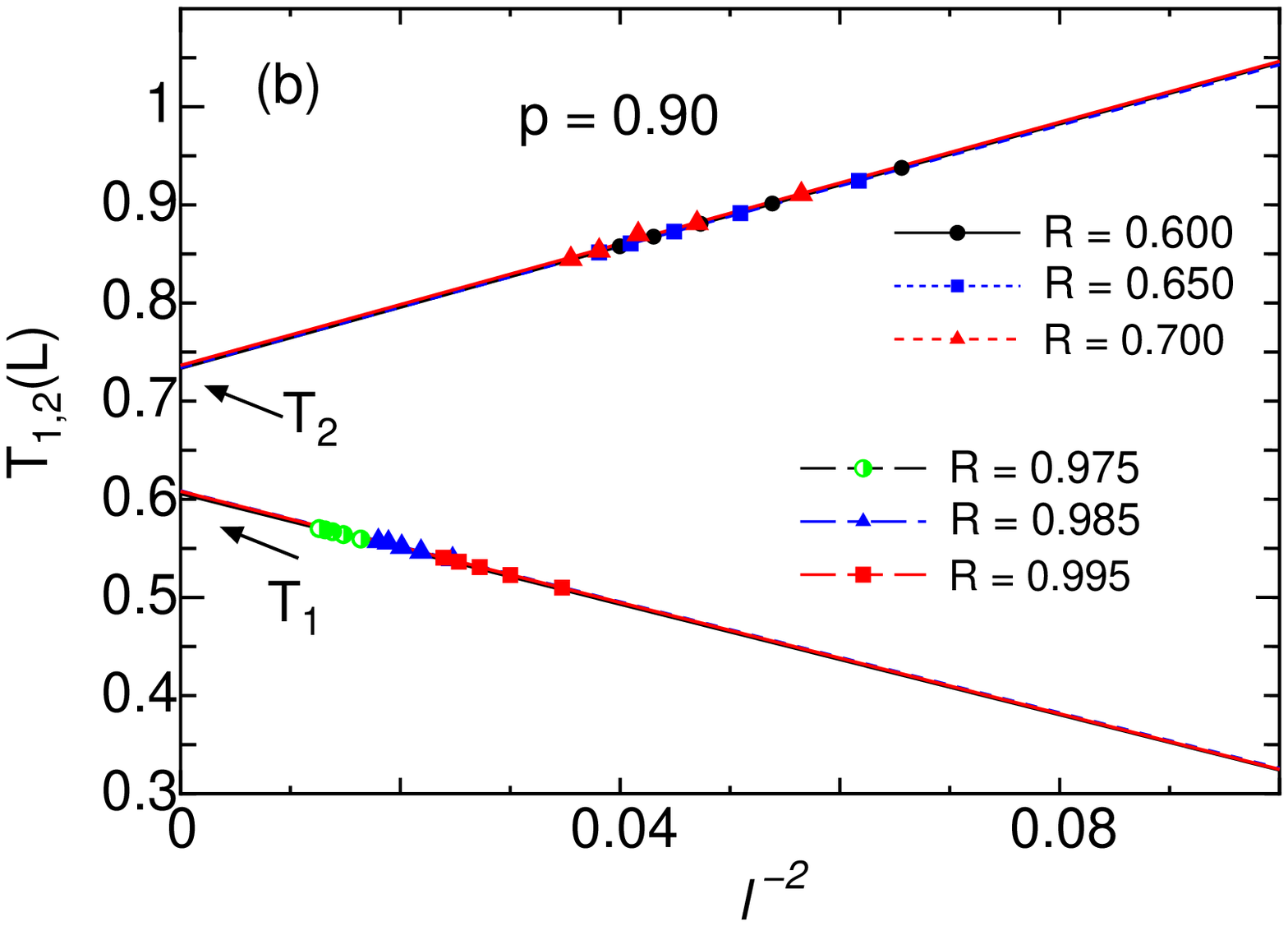}
\caption{The plot of $T_{KT}(L)$ versus $l^{-2}$, with $l = \ln bL$, 
to estimate KT transition temperature of the diluted (a) XY and (b) six-state clock models 
on the square lattice for $L$ = 32, 48, 64, 80, and 96, 
with bond concentration $p = 0.9$.  
For the clock model, both the higher ($T_2$) and lower ($T_1$) KT 
transition temperatures are estimated.  The data obtained from different values of $R$ 
are shown by different marks.}
\label{dilkt0.9} 
\end{center}
\end{figure}

\begin{figure}
\begin{center}
\includegraphics[width=0.9\linewidth]{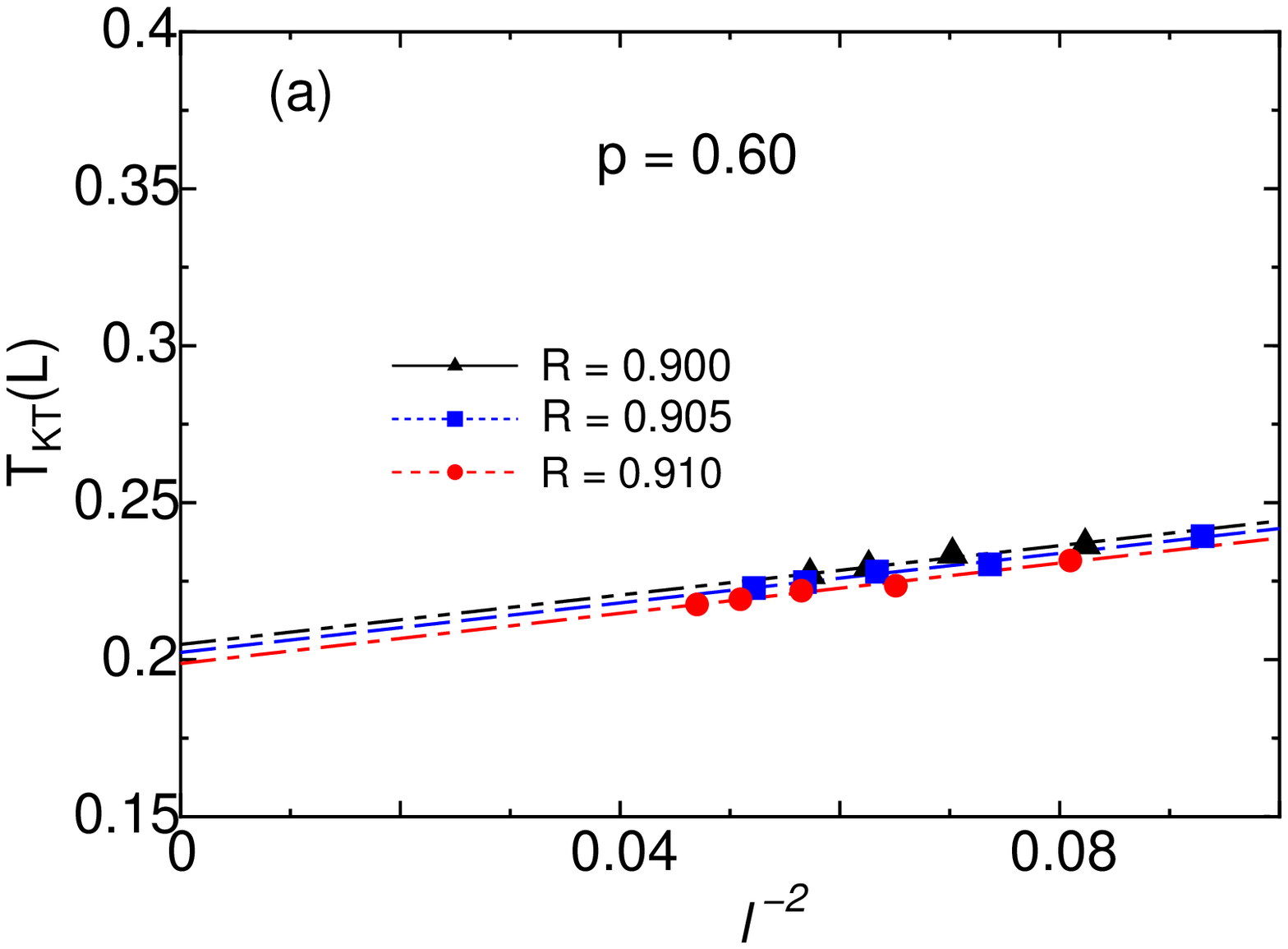}

\vspace{2mm}
\includegraphics[width=0.9\linewidth]{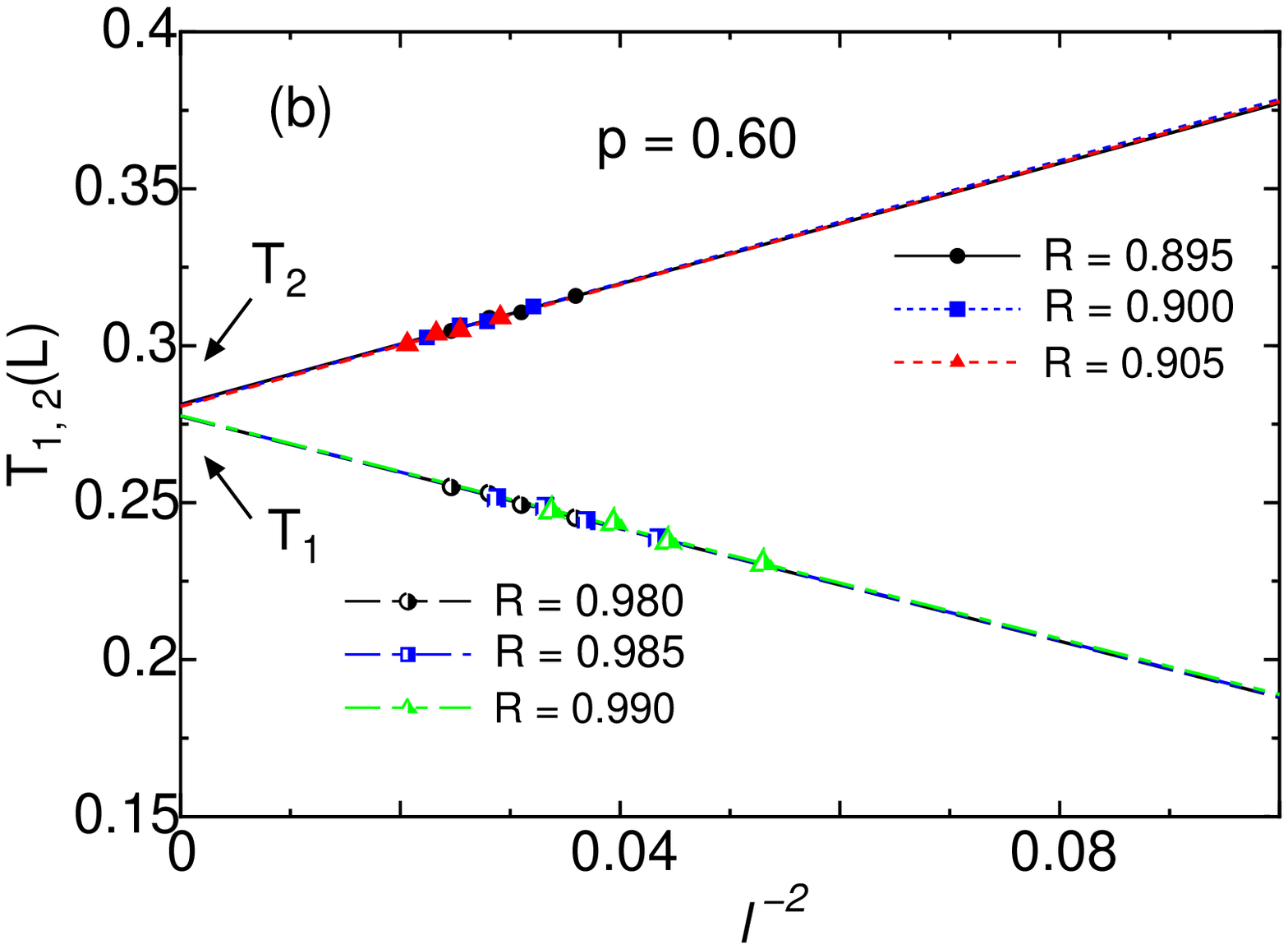}
\caption{The plot of $T_{KT}(L)$ versus $l^{-2}$, with $l = \ln bL$, 
to estimate KT transition temperature of the diluted (a) XY and (b) six-state clock models 
on the square lattice for $L$ = 32, 48, 64, 80, and 96, 
with bond concentration $p = 0.6$.  
For the clock model, both the higher ($T_2$) and lower ($T_1$) KT 
transition temperatures are estimated.  The data obtained from different values of $R$ 
are shown by different marks.}
\label{dilkt0.6} 
\end{center}
\end{figure}

We show the temperature dependence of magnetic correlation ratio 
for various bond concentrations $p$ in Fig.~\ref{ratioxy} 
for the diluted XY and six-state clock models. 
The data of different sizes for each bond concentration are plotted. 
The statistical errors are less than the order of the width of curves.  
Although at high temperatures, curves of different sizes are 
separated, they gradually merge as the temperature is decreased.  
At lower temperatures the curves of different sizes are collapsed 
on a single curve, which is the behavior of the QLRO phase, in other words, 
on the critical line. 
The essential feature of the diluted systems is the same 
as the pure case, which indicates the existence of the KT transition. 
The approach to QLRO phase is described by the scaling behavior 
shown in Eq.~(\ref{corr_ratio}). 
For the six-state clock model, the curves with different sizes 
separate again at low enough temperatures.  This comes from 
the discrete symmetry of the clock model, which yields the 
low-temperature ordered phase.  Thus, in addition to the 
high-temperature KT transition, a second KT transition 
at lower temperature exists for the six-state clock model. 
We find that the overall behavior of the diluted clock model is 
essentially the same as the pure clock model except that 
the KT transition temperatures decrease with the dilution. 
We shall estimate the KT transition temperatures of both the XY and 
clock models using FSS. 

With the FSS analysis based on Eq.~(\ref{corr_ratio}) and
the KT form of the correlation length, 
$\xi \propto \exp(c/\sqrt{t})$,
where $t=|T-T_{\rm KT}|/T_{\rm KT}$, we can write the $L$ dependence of $T_{\rm KT}(L)$ as follows 
\begin{equation}
 T_{\rm KT}(L) = T_{\rm KT} + \frac{c^2T_{\rm KT}}{(\ln bL)^2}. 
\label{T_KT}
\end{equation}
We consider the size-dependent critical temperature $T_{\rm KT}(L)$ 
that gives the constant $R=g(L/2)/g(L/4)$. 
In Fig.~\ref{dilkt0.9}, we show the plot of $T_{\rm KT}(L)$ 
as a function of $l^{-2}$, with $l=\ln (bL)$, 
using best-fitted parameters $b$ and $c$ for the diluted XY 
and six-state clock models with $p$ = 0.9. 
The system sizes are $L=$ 32, 48, 64, 80, and 96.
For the value of $R$, 0.600, 0.650, and 0.700 are used for the 
XY model, and $T_{\rm KT}(L)$ obtained using different $R$ are 
represented by different marks.  For the estimate of $T_2$ of 
the diluted clock model, the value of $R$ is set to be 0.600, 0.650, 
and 0.670, whereas $R$ is 0.975, 0.985, and 0.995 for $T_1$.  
The data using different $R$ are collapsed on a single curve,
which suggests that the difference of $R$ is absorbed 
in the $R$-dependence of $b$. 
The intercepts in the vertical axis of Fig.~\ref{dilkt0.9}
give the estimate of the KT transition temperatures. 
The estimate of $T_{\rm KT}$ for the diluted XY model with $p=0.9$ 
is 0.747(4).  The number in the parentheses denotes the uncertainty 
in the last digits.  In the same way, we estimate the two KT transition 
temperatures of the diluted six-state clock model, $T_2$ and $T_1$, as 
0.727(4) and 0.605(4), respectively. 

We also plot the data for $p=0.6$ in Fig.~\ref{dilkt0.6}. 
The choice of fixed $R$ is different from that for $p=0.9$. 
The estimated KT transition temperatures are given by the intercepts 
in the vertical axis.  The KT transition temperatures for $p=0.6$ are 
lower than those for $p=0.9$; for the six-state clock model, 
the QLRO phase between $T_1$ and $T_2$ becomes narrower. 

\begin{table}
\caption{The estimates of KT transition temperatures for the diluted XY and 
six-state clock models for various bond concentration $p$. 
For the clock model the lower ($T_1$) and higher ($T_2$) KT transition 
temperatures are estimated.}
\label{diltable1}
\vspace{2mm}
\begin{ruledtabular}
\begin {tabular}  {llll}
~~~~~~~~~~~ & XY~~~~~~~~~~~ & six-state clock
\\
~$p$ & ~$T_{\rm{KT}}$ & ~$T_1$ & ~$T_2$ \\
\tableline
1.0~ & 0.895(3)~ & 0.715(3)~ & 0.902(3) \\
0.9  & 0.747(4)  & 0.605(4)  & 0.727(4) \\
0.8  & 0.575(6)  & 0.489(6)  & 0.574(4) \\
0.7  & 0.401(6)  & 0.388(6)  & 0.434(6) \\
0.6  & 0.215(12) & 0.277(4)  & 0.281(4) \\
0.55 & 0.076(8)  & 0.198(8)  & 0.204(6) \\
\end{tabular}
\end{ruledtabular}
\end{table}

\begin{figure}
\begin{center}
\includegraphics[width=0.9\linewidth]{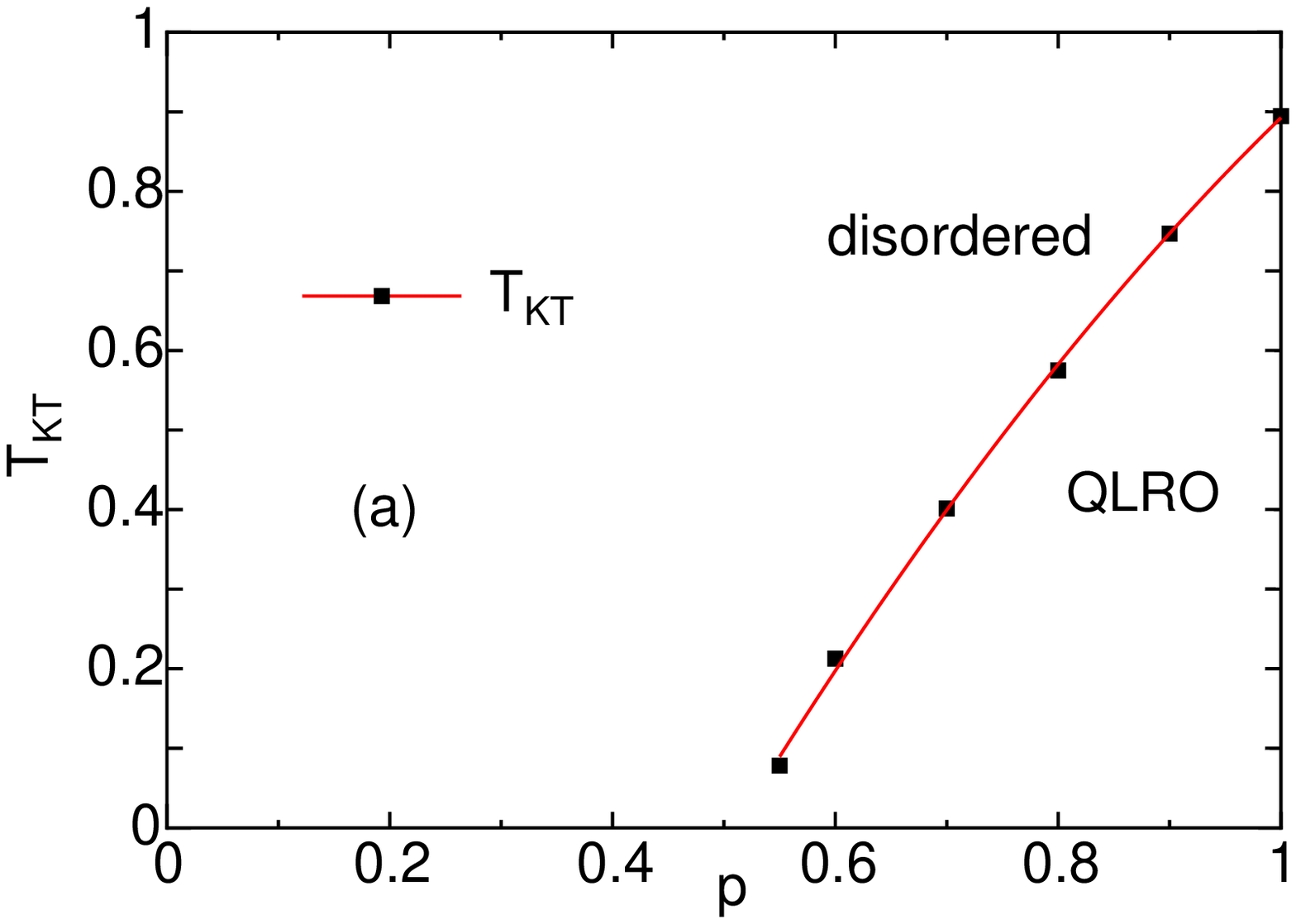}

\vspace{2mm}
\includegraphics[width=0.9\linewidth]{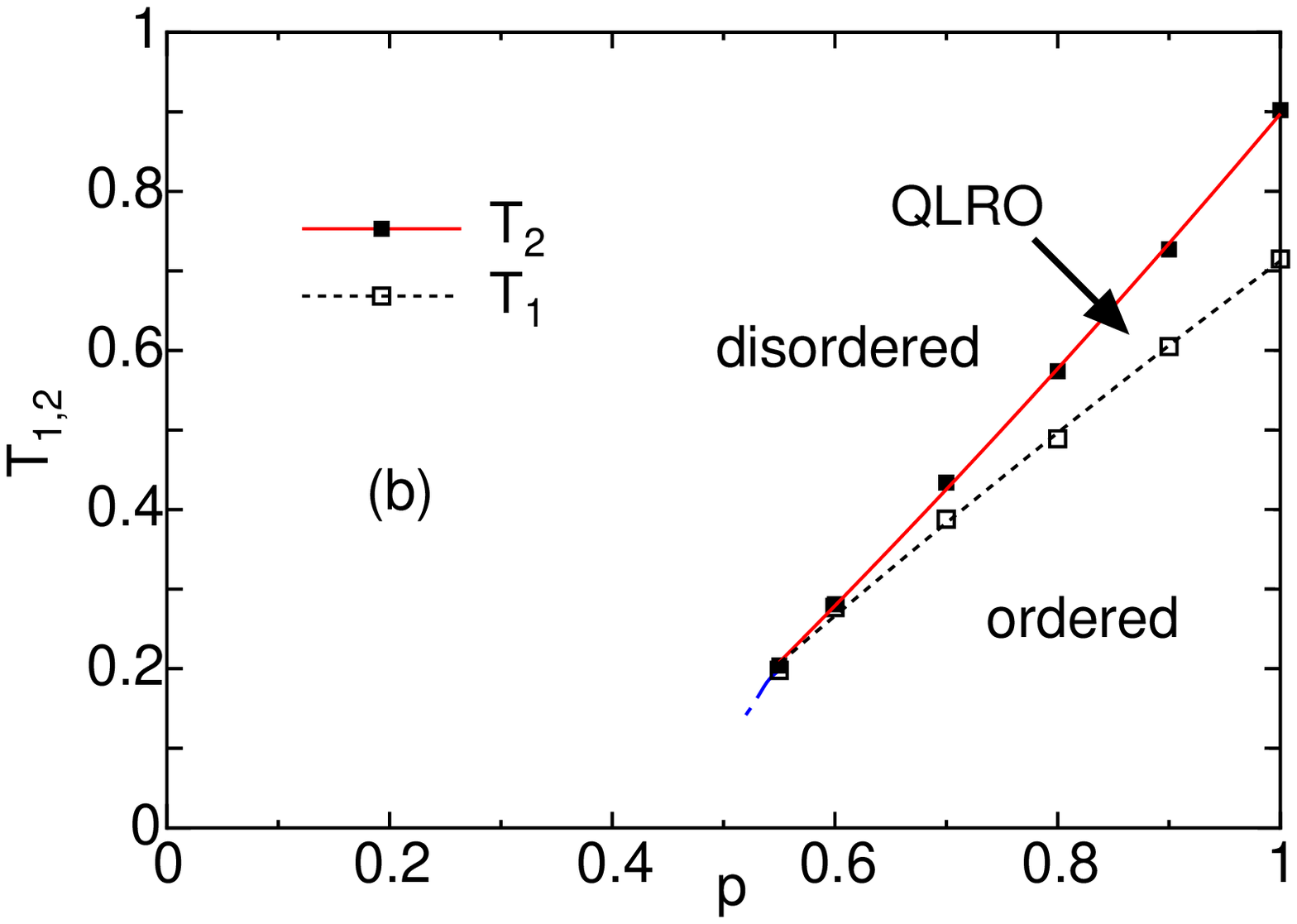}
\caption{Phase diagram of the diluted (a) XY and (b) six-state 
clock models.  As shown, there is a systematic shift of 
the KT transition temperatures as the bond concentration decreases.  
The plot suggests that the critical dilution is close to 
the percolation threshold $p_c = 0.5$.  
For the clock model, due to the discrete symmetry, 
there exists a lower KT transition, 
separating the ordered phase of low temperature 
from the intermediate QLRO.}
\label{dilphase} 
\end{center}
\end{figure}

Performing the same procedure for other bond concentrations,  
we estimate their KT transition temperatures; they are tabulated 
in Table \ref{diltable1}. For the six-state clock model, 
the lower ($T_1$) and higher ($T_2$) KT transition 
temperatures are estimated. 
In Fig.~\ref{dilphase} we show the phase diagram of the diluted systems, 
which is produced from Table \ref{diltable1}. 
As can be seen from the phase diagram, the KT transition temperatures 
gradually decrease with dilution, and the QLRO phase 
disappears at the bond concentration which is very close to 
the percolation threshold $p_c$;
for the bond percolation of the square lattice, $p_c=0.5$. 
This result is the same as that by Berche {\it et al.}~\cite{berche} 
for the site dilution, 
but different from that by Leonel {\it et al.}~\cite{leonel}.  

For the clock model the intermediate QLRO phase is suppressed to 
a narrow range of temperature as the diluted bonds increase.  
The lower KT transition comes from the discrete symmetry.  
Thus, the behavior near the percolation threshold is similar to the 
case of the diluted Ising model.  
The higher KT transition temperature $T_2$ for the clock model becomes 
higher than $T_{\rm KT}$ for the XY model with the same $p$ near $p=0.5$, 
which may be related to the stabilization effect of discrete symmetry. 

\subsubsection{Decay exponent}

\begin{figure}
\begin{center}
\includegraphics[width=0.9\linewidth]{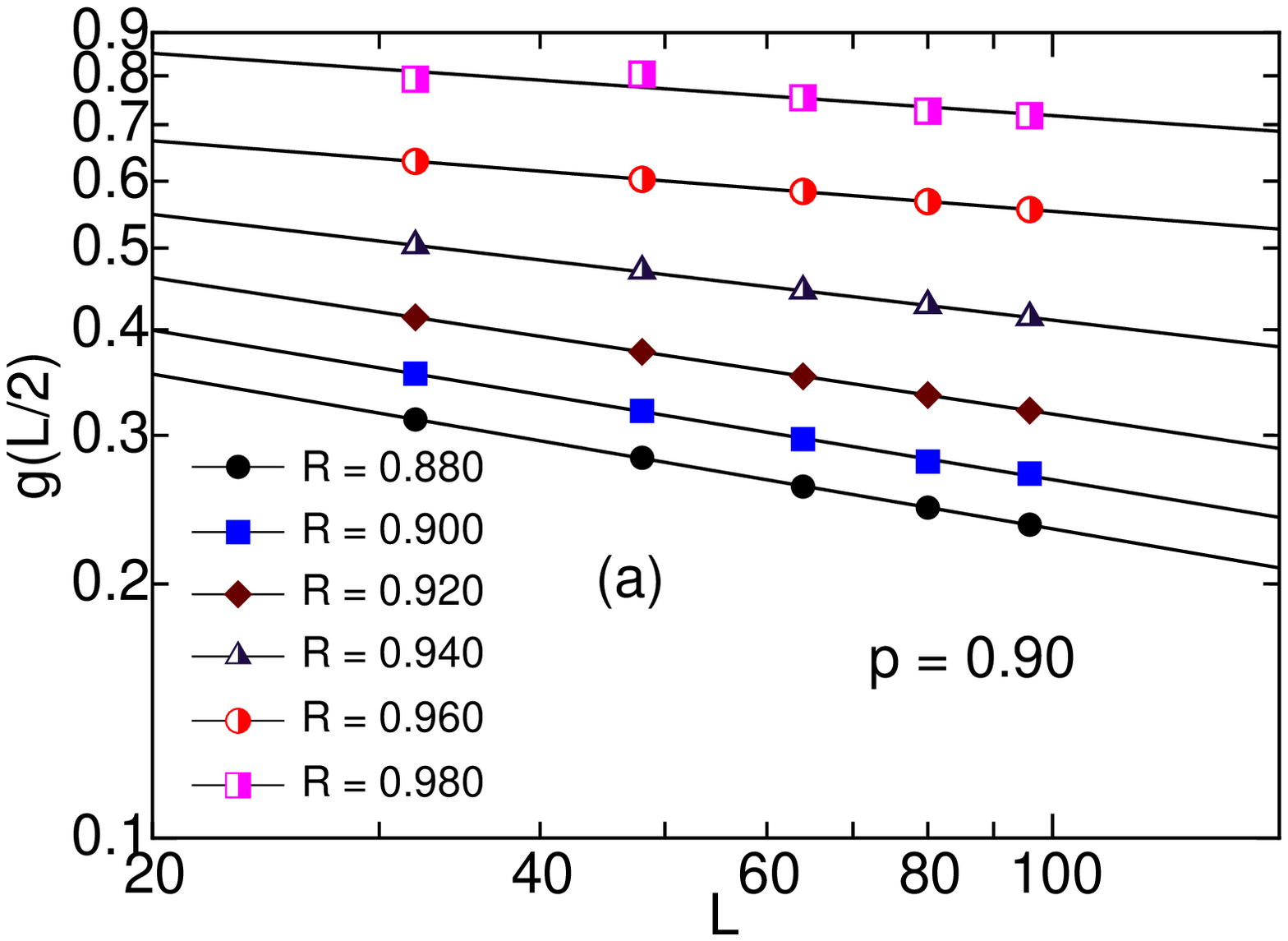}
\includegraphics[width=0.9\linewidth]{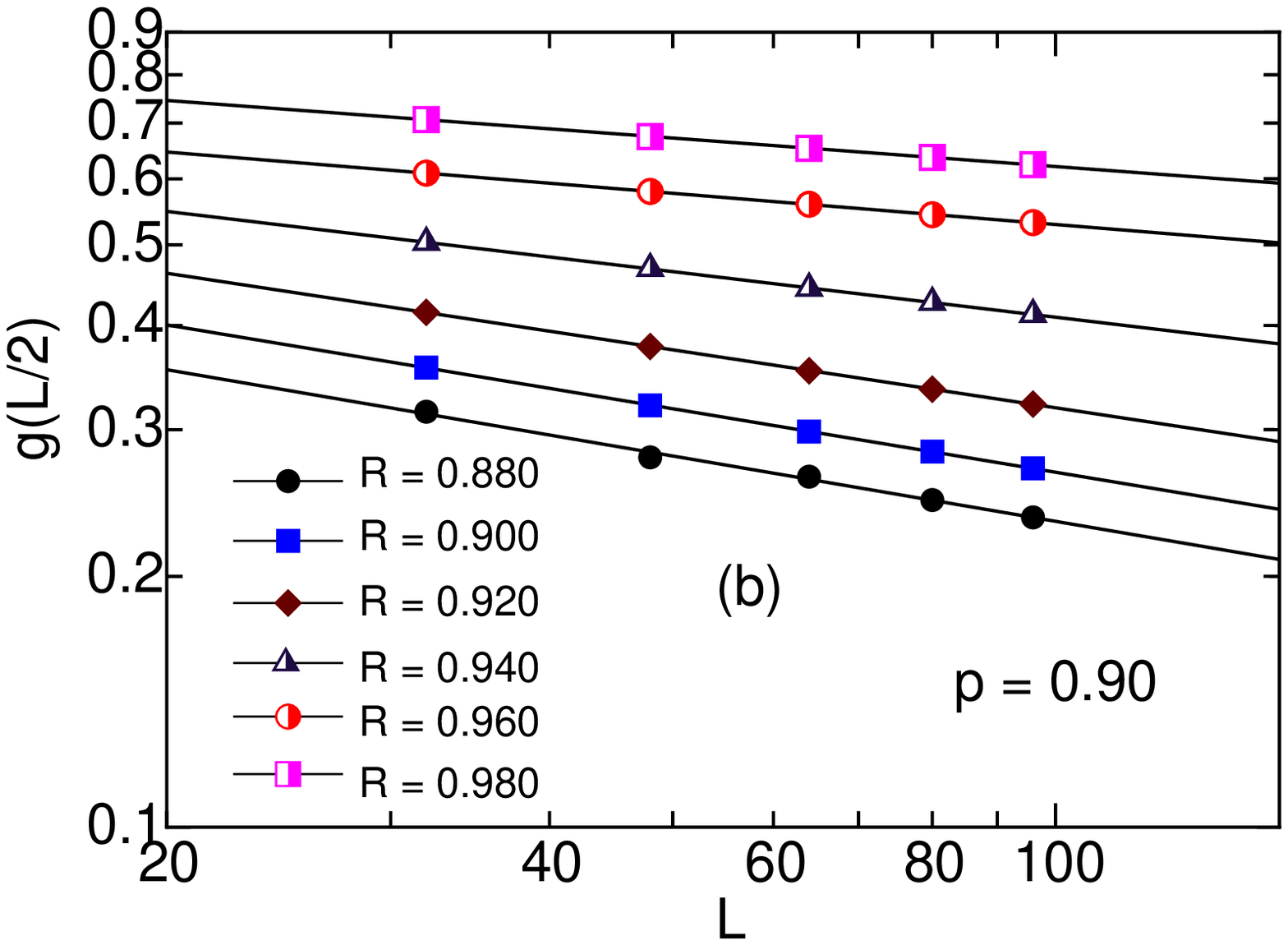}
\caption{Double-logarithmic plot of the magnetic correlation function 
$g(L/2)$ versus $L$ of the diluted (a) XY and (b) six-state clock models, 
for bond concentration $p = 0.9$. Here the slope of 
the best-fitted line of each corresponding $R$ gives the estimate  
of the exponent $\eta$.}
\label{g2_0.9} 
\end{center}
\end{figure}

\begin{figure}
\begin{center}
\includegraphics[width=0.9\linewidth]{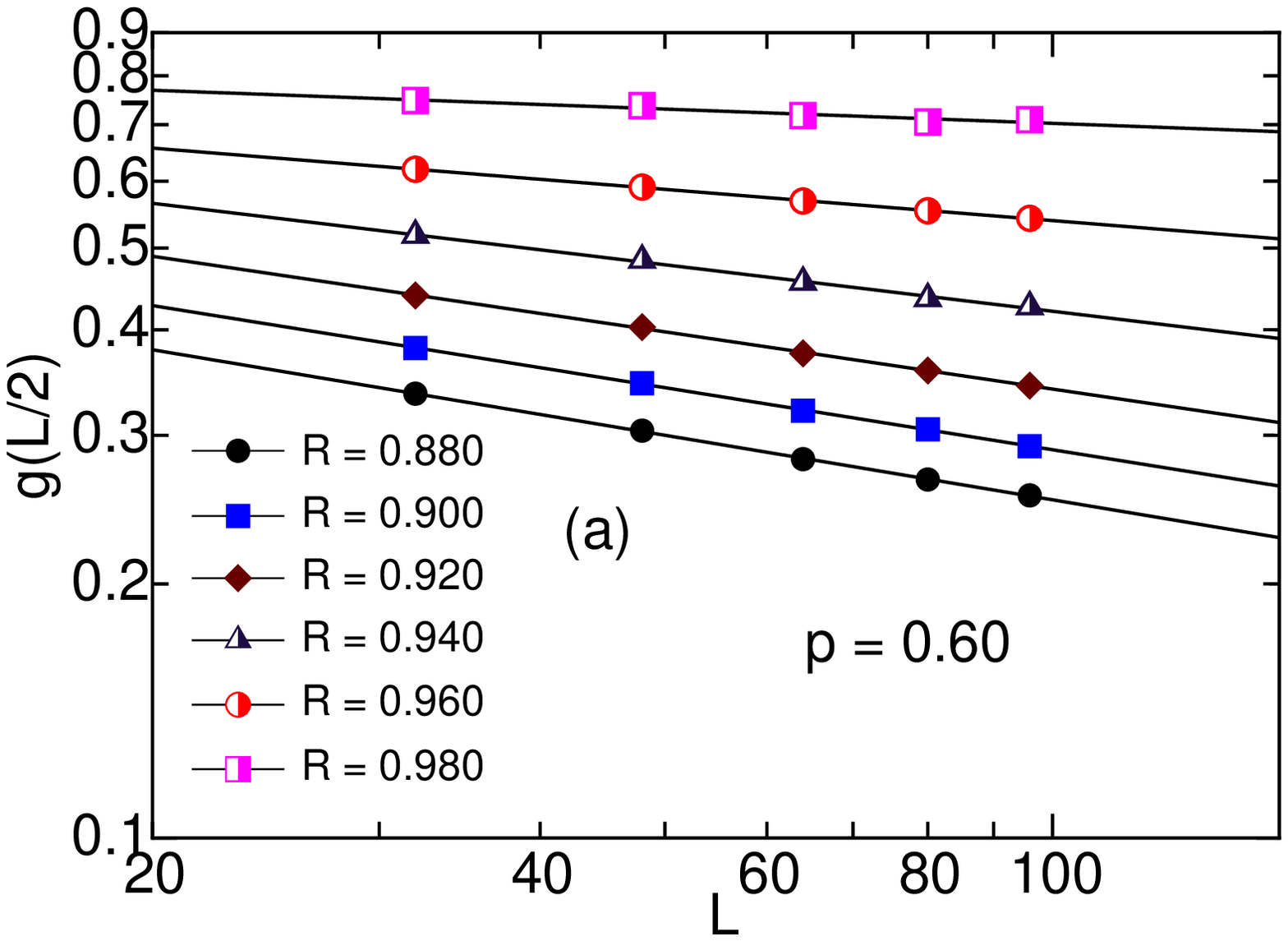}
\includegraphics[width=0.9\linewidth]{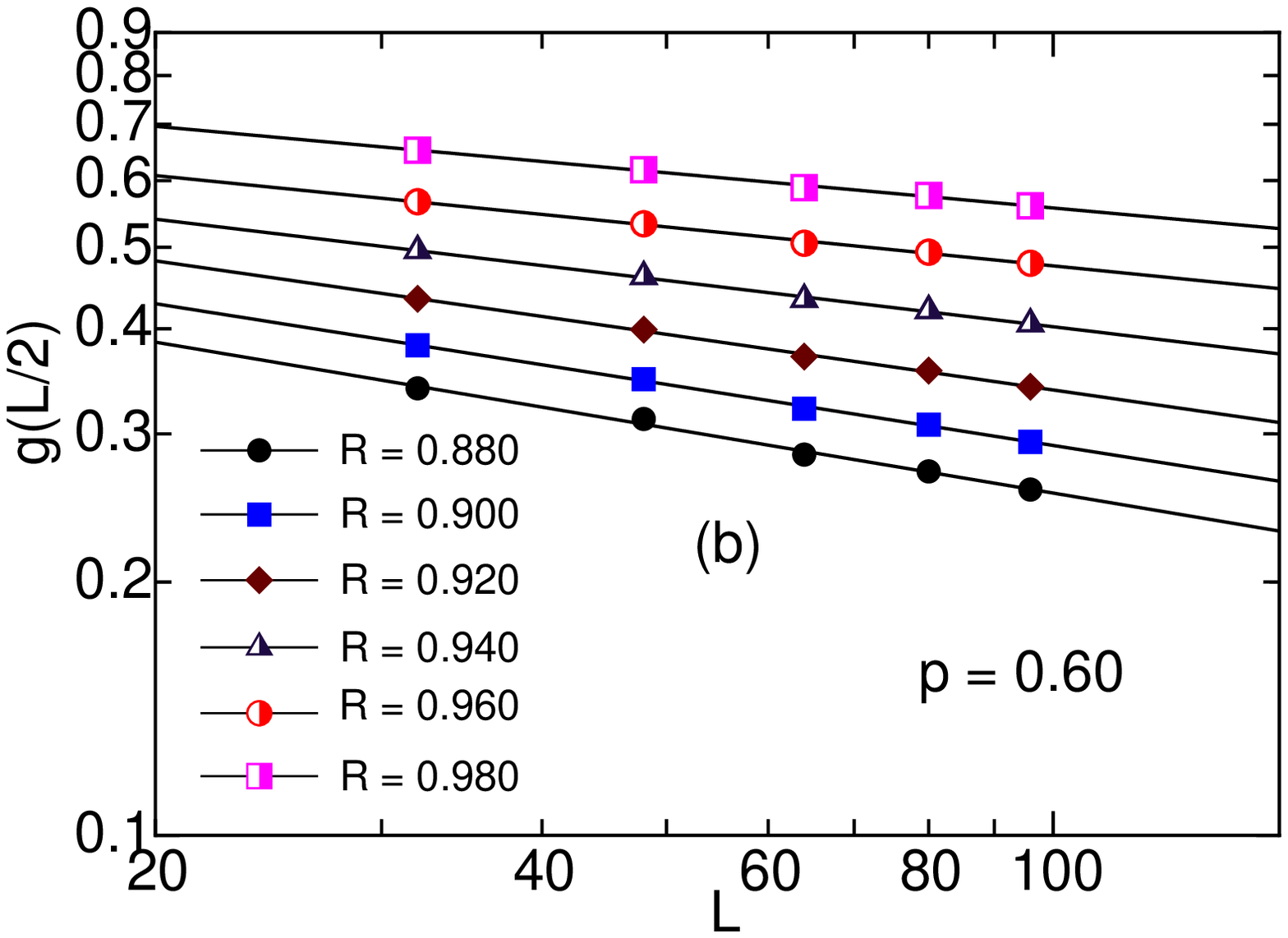}
\caption{Double-logarithmic plot of the magnetic correlation function 
$g(L/2)$ versus $L$ of the diluted (a) XY and (b) six-state clock models, 
for bond concentration $p = 0.6$. Here the slope of 
the best-fitted line of each corresponding $R$ gives the estimate  
of the exponent $\eta$.}
\label{g2_0.6} 
\end{center}
\end{figure}

Next, we consider the decay exponent $\eta$ of both XY and six-state 
clock models for each bond concentration.  We first look at 
the constant value of correlation ratio $R$ for different sizes 
and find the associate correlation function $g(L/2)$. 
We give attention to the power-law dependence of the correlation function 
on the system size, $g(L/2) \sim L^{-\eta}$, expressed in Eq.~(\ref{g(r)}) 
with $D = 2$.  Choosing the fixed $R$, we have the same 
temperature for different sizes on the critical point or on the critical 
line.  Moreover, away from the critical points the same $R$ gives 
different temperatures for different sizes, but the corrections 
to the power-law behavior, Eq.~(\ref{g(r)}), is the same, 
which yields the estimate of the decay exponent $\eta$.  
This analysis of $\eta$ was used in the study of the fully 
frustrated clock model~\cite{tasrief}. 
As an example, we consider the system with bond concentration $p = 0.9$. 
We plot $g(L/2)$ versus $L$ for various $R$'s in double-logarithmic scale 
for the diluted XY and six-state clock models in Fig.~\ref{g2_0.9}.  
The value of $\eta$ is estimated from the slope of the best-fitted line 
for each value of constant correlation ratio.   
Similar plots for $p=0.6$ are given in Fig.~\ref{g2_0.6}.  

\begin{figure}
\begin{center}
\includegraphics[width=0.9\linewidth]{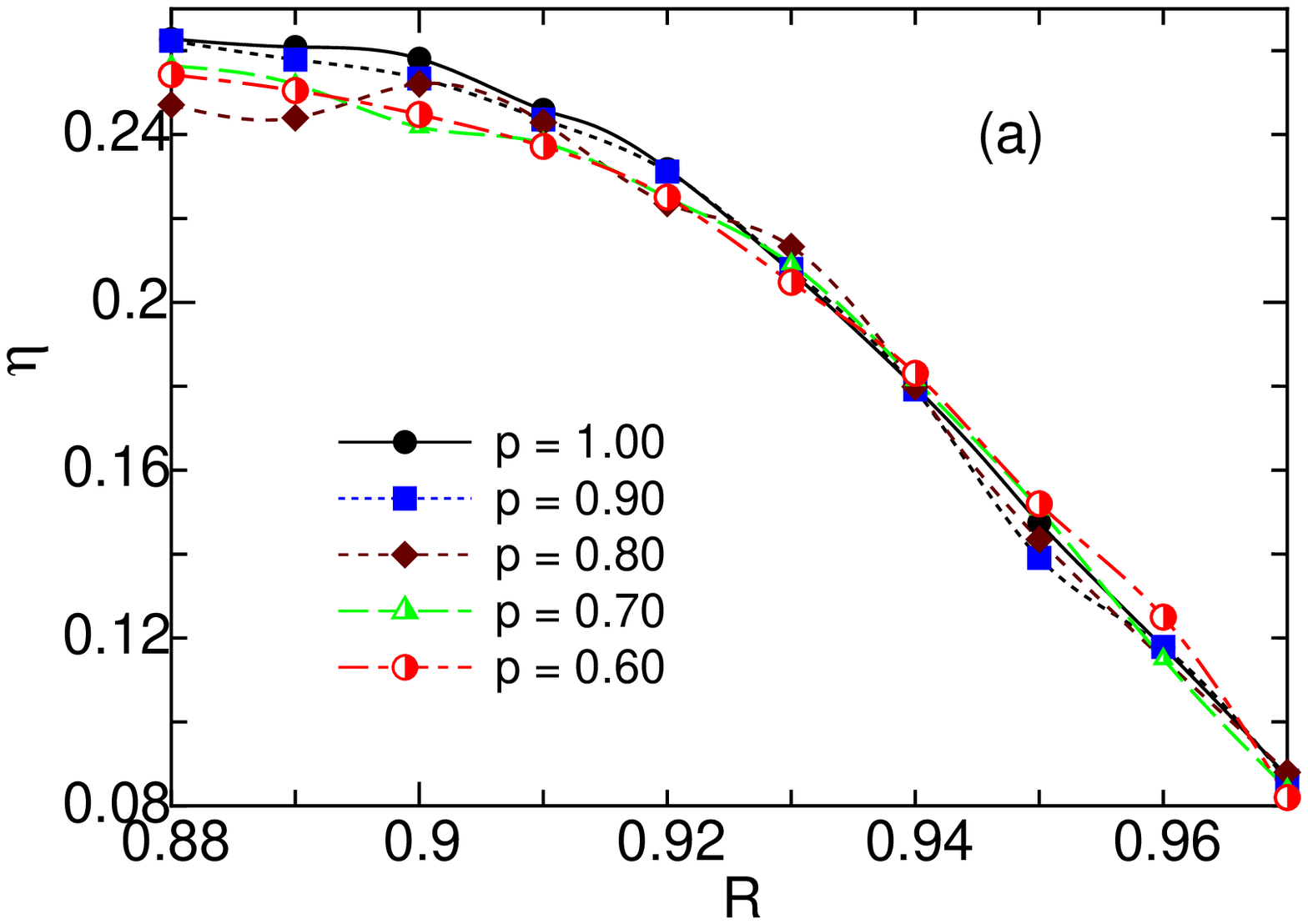}

\vspace{2mm}
\includegraphics[width=0.9\linewidth]{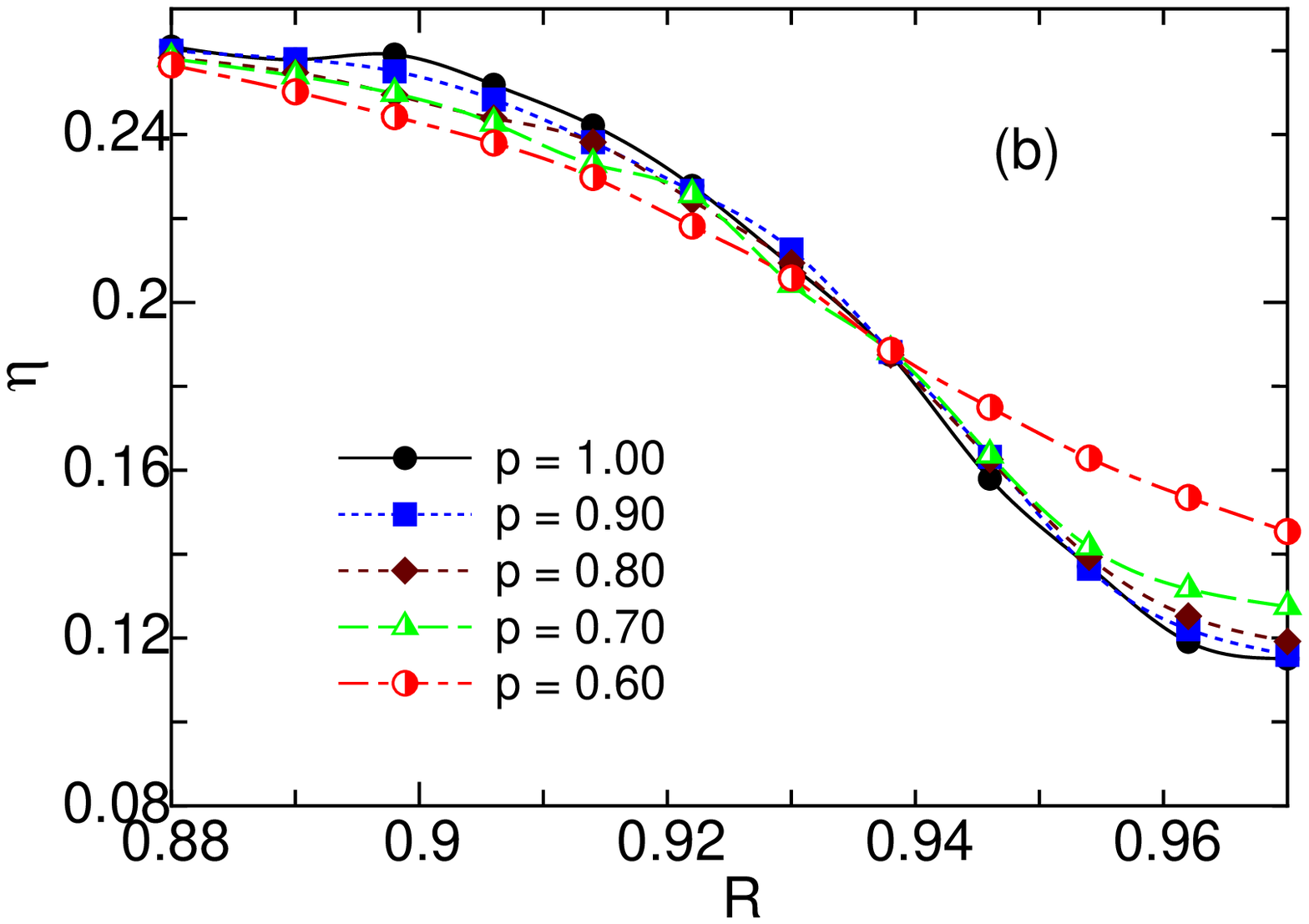}
\caption{Decay exponent $\eta$ as function of correlation ratio $R$ 
of the diluted (a) XY and (b) six-state clock models 
for various concentrations. }
\label{dileta}
\end{center}
\end{figure}

The $R$-dependence of thus obtained $\eta$ of each bond concentration 
for the diluted XY and six-state clock models are 
plotted in Fig.~\ref{dileta}. 
As can be seen, the exponents $\eta$ of bond diluted system with various 
bond concentrations behave similar to those of pure case, $p=1$, 
namely continuously changing with the temperature in the KT phase. 
We have plotted the data down to $p=0.6$.  All the data seem 
to be universal, and the corrections are small except for larger $R$ 
side of the clock model with $p=0.6$.  The deviations from the pure 
value becomes larger for $p = 0.55$; they are not plotted here.  
This comes from the fact that 
it is close to the percolation threshold and corrections become larger. 
In the renormalization-group language, the critical behavior 
is affected by another fixed point nearby. 

Since the estimated $\eta$ is almost constant for smaller $R$, 
which corresponds to higher temperature, in Fig.~\ref{dileta}, 
the exponent at $T_{\rm KT}$ of the XY model and $T_2$ of the clock model
are estimated as 
$$
   \eta_2 = 0.25(1).  
$$
This value is consistent with that for the pure case, 1/4=0.25. 
For the low-temperature side (large $R$) of the XY model, 
$\eta$ approaches 0.  
Meanwhile, the estimated $\eta$ for the six-state clock model 
becomes constant for large $R$.  This gives the decay exponent 
at the lower KT transition temperature $T_1$ as
$$
\eta_1 = 0.12(1), 
$$
which is consistent with the pure value of $1/9=0.111$. 
The present results suggest that the exponents associated 
with the KT transitions are universal with dilution.

\section{Concluding Remarks}\label{conclu}

In summary, we have investigated the bond dilution of the XY and 
six-state clock models on the square lattice 
using the canonical sampling MC method with cluster spin updates.  
We have observed the KT transition of higher temperature 
separating the intermediate QLRO phase from the disordered phase. 
Due to the discrete symmetry of the clock model, 
the lower KT transition has been also observed. 
There is a systematic decrease in KT transition temperatures 
as the concentration of diluted bonds increases. 
Our estimates of the KT transition temperatures for each concentration, 
both for the XY and six-state clock models, are listed 
in Table \ref{diltable1}, 
from which the corresponding phase diagram shown 
in Fig.~\ref{dilphase} follows. 
As can be seen from the phase diagram, the critical concentration 
of dilution is very close to the percolation threshold 
which is the same as the result by Berche {\it et al.}~\cite{berche} 
for the site dilution, but different from that 
by Leonel {\it et al.}~\cite{leonel}.  
The bond-diluted and site-diluted classical spin systems show 
essentially the same behavior, although there may be differences 
for the quantum spin models~\cite{yasuda1,yasuda2}. 
Our prelimenary results for the site-diluted XY and clock models 
give the continuous phase transition with respect to $p$.  
Thus, our result is in favor of that by Berche {\it et al.}~\cite{berche}.

The phase diagram also shows that 
the intermediate QLRO phase for the clock model is suppressed 
to a narrow range of temperature as the diluted bonds increase. 
Our estimates of decay exponents for lower concentration dilution 
suggest that the KT transition remains unaffected by 
dilution, which is analogous to the expectation of the Harris criterion 
that the randomness is irrelevant for system with non-diverging specific heat.

Quite recently, Sasamoto and Nishimori have studied the phase diagram 
of 2D random $Z_q$ models using the duality argument~\cite{sasamoto}. 
The analysis of the duality argument for the present model will be 
left to a future problem.

\section*{Acknowledgments}

The authors wish to thank N. Kawashima, H. Otsuka, C. Yamaguchi, 
Y. Tomita, M. Ito, H. Nishimori and T. Sasamoto for valuable discussions.  
One of the authors (TS) gratefully acknowledges the fellowship 
provided by the Ministry of Education, Science, Sports and Culture, Japan. 
This work was supported by a Grant-in-Aid for Scientific Research 
from the Japan Society for the Promotion of Science.
The computation of this work has been done using computer facilities 
of Tokyo Metropolitan University and those of 
the Supercomputer Center, Institute for Solid State Physics, 
University of Tokyo.

\end{document}